\documentclass[runningheads,11pt]{llncs}
\usepackage{graphicx}
\usepackage{fullpage}

\usepackage{comment}
\usepackage{amsmath}
\usepackage{url}
\usepackage{multirow}

\def\SA{\mbox{\rm {\sf SA}}}
\def\ISA{\mbox{\rm {\sf ISA}}}

\def\lcp{\mbox{\rm {\sf lcp}}}

\def\X{\mbox{\rm {\sf X}}}

\def\PSVL{\mbox{$\text{\sf PSV}_{\text{\sf lex}}$}}
\def\PSVT{\mbox{$\text{\sf PSV}_{\text{\sf text}}$}}
\def\NSVL{\mbox{$\text{\sf NSV}_{\text{\sf lex}}$}}
\def\NSVT{\mbox{$\text{\sf NSV}_{\text{\sf text}}$}}

\def\O{\mbox{\rm O}}
\DeclareMathSymbol{\Theta}{\mathalpha}{operators}{2}
\DeclareMathSymbol{\Phi}{\mathalpha}{operators}{8}

\def\balgorithm#1{{\bf Algorithm \texttt{#1}}}

\def\bfor{{\bf for\ }}

\def\bto{{\bf to\ }}

\def\bwhile{{\bf while\ }}

\def\bdo{{\bf do\ }}
\def\bif{{\bf if\ }}
\def\bthen{{\bf then\ }}
\def\belse{{\bf else\ }}

\def\breturn{{\bf return\ }}
\def\boutput{{\bf output\ }}

\def\la{\leftarrow}

\def\+{\!+\!}
\def\-{\!-\!}

\begin{document}
\title{Linear Time Lempel-Ziv Factorization: Simple, Fast, Small}

\author{
Juha K{\"a}rkk{\"a}inen
\and
Dominik Kempa
\and
Simon J. Puglisi
}

\institute{
    Department of Computer Science,\\
    University of Helsinki\\
    Helsinki, Finland\\
    \email{firstname.lastname@cs.helsinki.fi}\\[1ex]
}

\date{}

\maketitle \thispagestyle{empty}

\begin{abstract}
  Computing the LZ factorization (or LZ77 parsing) of a string is a
  computational bottleneck in many diverse applications, including
  data compression, text indexing, and pattern discovery. We describe
  new linear time LZ factorization algorithms, some of which require 
  only
  $2n\log n + \O(\log n)$ bits of working space to factorize a string of
  length $n$. 
  These are the most space efficient linear time algorithms
  to date, using $n \log n$ bits less space
than any previous linear time algorithm.
The algorithms are also practical, simple to implement, and  
very fast in practice. 
\end{abstract}

\section{Introduction} \label{sec:intro} In the 35 years since its
discovery the LZ77 factorization of a string --- named after its
authors Abraham Lempel and Jacob Ziv, and the year $1977$ in which it
was published --- has been applied all over computer science. The
first uses of LZ77 were in data compression, and to this day it lies
are the heart of efficient and widely used file compressors, like {\tt
  gzip} and {\tt 7zip}. LZ77 is also important as a {\em measure} of
compressibility.  For example, its size is a lower bound on the size
of the smallest context-free grammar that represents a
string~\cite{cllppss2005}.

In all these applications (and most of the many others we have not
listed) computation of the factorization is a time- and
space-bottleneck in practice. Our particular motivation is the
construction of compressed full-text indexes~\cite{nm2007}, several
recent and powerful instances of which are based on
LZ77~\cite{ggp2011,ggknp2012,kn2011}.

\paragraph{Related work.}
There exists a variety of worstcase linear time algorithms to compute
the LZ factorization~\cite{cps07,ci2008,cis2008,gb2012,og2011}. 
All of them
require at least $3n \log n$ bits of working space\footnote{The
  working space excludes the input string, the output factorization,
  and $\O(\log n)$ terms.} 
in the worstcase. 
The most space efficient linear time algorithm is due to Chen
et al.~\cite{cps07}. By overwriting the suffix array it achieves a
working space of $(2n+s) \log n$ bits, where $s$ is the maximal size
of the stack used in the algorithm. However, in the worstcase
$s=\Theta(n)$. Another space efficient solution requiring
$(2n+\sqrt{n}) \log n$ bits of space in the worstcase is
from~\cite{cis2008} but it computes only the lengths of LZ77
factorization phrases. It can be extended to compute the full parsing
at the cost of extra $n \log n$ bits.

All of these algorithms rely on the suffix
array, which 
can be constructed in $\O(n)$ time and using $(1+\epsilon)n \log n$ bits of
space (in addition to the input string but including the output of size $n
\log n$ bits)~\cite{ksb2006}. 
This raises the question of whether the space complexity of
linear time LZ77 factorization can be reduced from $3n \log n$ bits.
In this paper, we answer the question in the affirmative by describing
a linear time algorithm using $2n\log n$ bits. 

In terms of practical performance, the fastest linear time LZ
factorization algorithms are the very recent ones by Goto and
Bannai~\cite{gb2012}, all using at least $3n\log n$ bits of working
space.  Other candidates for the fastest algorithms are described by
Kempa and Puglisi~\cite{kp2013}.  
Due to nearly simultaneous publication, 
no comparison between them exists so far.
Experiments in this paper put the algorithms of Kempa and Puglisi
slightly ahead. Their algorithms are also very space efficient; one of
them uses $2n\log n + n$ bits of working space and others even less.
However, their worstcase time complexity is $\Theta(n\log\sigma)$ for an
alphabet of size $\sigma$.
More details about these algorithms are given in
Section~\ref{sec-preliminaries}. 

\paragraph{Our contribution.} 
We describe two linear time algorithms for LZ factorization.
The first algorithm uses $3n \log n$ bits of working space and can be
seen as a reorganization of an algorithm by Goto and
Bannai~\cite{gb2012}. However, this reorganization makes it smaller
and faster. In our experiments, this is the fastest of all
algorithms when the input is not highly repetitive. 

The second algorithm employs a novel combinatorial technique to reduce
the working space to $2n \log n$ bits, which is at least $n\log n$
bits less than any previous linear time algorithm uses in the
worstcase. The space reduction does not come at a great cost in
performance. The algorithm is the fastest on some inputs and not far
behind the fastest on others.

Both algorithms share several nice features. They are
alphabet-independent, using only character comparisons to access the
input. They make just one sequential pass over the suffix array,
enabling streaming from disk, which would reduce the working space
by a further $n\log n$ bits.  They are also very simple and easy to
implement. 

\section{Preliminaries}
\label{sec-preliminaries}

\paragraph{Strings.}
Throughout we consider a string $\X = \X[1..n] = \X[1]\X[2]\ldots
\X[n]$ of $|\X| = n$ symbols
drawn from an ordered alphabet of size $\sigma$.

For $i=1,\ldots,n$ we write $\X[i..n]$ to denote the {\em suffix} of
$\X$ of length $n-i+1$, that is $\X[i..n] = \X[i]\X[i+1]\ldots \X[n]$.
We will often refer to suffix $\X[i..n]$ simply as ``suffix
$i$''. Similarly, we write $\X[1..i]$ to denote the {\em prefix} of
$\X$ of length $i$.  We write $\X[i..j]$ to represent the {\em
  substring} $\X[i]\X[i+1]\ldots \X[j]$ of $\X$ that starts at
position $i$ and ends at position $j$.
Let $\lcp(i,j)$ denote the length of the longest-common-prefix of
suffix $i$ and suffix $j$. For example, in the string $\X =
zzzzzipzip$, $\lcp(2,5) = 1 = |z|$, and $\lcp(5,8) = 3 = |zip|$. For
technical reasons we define $\lcp(i,0)=\lcp(0,i)=0$ for all $i$.

\paragraph{Suffix Arrays.} 
The suffix array
$\SA$ is an array $\SA[1..n]$ containing a permutation of the
integers $1..n$ such that $\X[\SA[1]..n] < \X[\SA[2]..n] < \cdots <
\X[\SA[n]..n]$.  In other words, $\SA[j] = i$ iff $\X[i..n]$ is the
$j^{\mbox{{\scriptsize th}}}$ suffix of $\X$ in ascending
lexicographical order. 
The inverse suffix array $\ISA$ is the inverse
permutation of $\SA$, that is $\ISA[i] = j$ iff $\SA[j] =
i$. Conceptually, $\ISA[i]$ tells us the position of suffix $i$ in
$\SA$. 

The array $\Phi[0..n]$ (see~\cite{KarkkainenMP09}) is defined by $\Phi[i]=\SA[\ISA[i]-1]$, that is,
the suffix $\Phi[i]$ is the immediate lexicographical predecessor of
the suffix $i$. For completeness and for technical reasons we define
$\Phi[\SA[1]]=0$ and $\Phi[0]=\SA[n]$ so that $\Phi$ forms a
permutation with one cycle.

\paragraph{LZ77.} 
The LZ77 factorization uses the notion of a {\em longest previous
  factor} (LPF).  The LPF at position $i$ in $\X$ is a pair
$(p_i,\ell_i)$ such that, $p_i < i$, $\X[p_i..p_i+\ell_i-1] =
\X[i..i+\ell_i-1]$ and $\ell_i>0$ is maximized.
In other
words, $\X[i..i+\ell_i-1]$ is the longest prefix of $\X[i..n]$ which
also occurs at some position $p_i < i$ in $\X$. If $\X[i]$
is the leftmost occurrence of a symbol in $\X$ then such a pair 
does not
exist. In this case we define 
$p_i = \X[i]$ and
$\ell_i = 0$. Note that there may be more than one potential
$p_i$, and we do not care which one is used.

The LZ77 factorization (or LZ77 parsing) of a string $\X$ is then just
a greedy, left-to-right parsing of $\X$ into longest previous
factors. More precisely, if the $j$th LZ factor (or {\em phrase}) in
the parsing is to start at position $i$, then we output $(p_i,\ell_i)$
(to represent the $j$th phrase), and then the $(j+1)$th phrase starts
at position $i+\ell_i$, unless $\ell_i = 0$, in which case the next
phrase starts at position $i+1$.  
We call a factor $(p_i,\ell_i)$ {\em
  normal} if it satisfies $l_i>0$ and {\em special} otherwise.
The number of phrases in the factorization is denoted by $z$.

For the example string $\X = zzzzzipzip$, the LZ77 factorization
produces:
$$(z,0),(1,4),(i,0),(p,0),(5,3).$$
The second and fifth factors are normal, and the other three are special.

\paragraph{NSV/PSV.} The LPF pairs can be computed
using \emph{next and previous smaller values} (NSV/PSV) defined as
\begin{align*}
  \NSVL[i] &= \min \{ j\in [i+1..n] \mid \SA[j] < \SA[i]\} \\
  \PSVL[i] &= \max \{ j\in [1..i-1] \mid \SA[j] < \SA[i]\} .
\end{align*}
If the set on the right hand side is empty, we set the value to $0$.
Further define
\begin{align}
  \NSVT[i] &= \SA[\NSVL[\ISA[i]]] \label{eqn-NSVT}\\
  \PSVT[i] &= \SA[\PSVL[\ISA[i]]] . \label{eqn-PSVT}
\end{align}
If $\NSVL[\ISA[i]]=0$ ($\PSVL[\ISA[i]]=0$) we set $\NSVT[i]=0$
($\PSVT[i]=0$). 

If $(p_i,\ell_i)$ is a normal factor, then either $p_i=\NSVT[i]$ or
$p_i=\PSVT[i]$ is always a valid choice for $p_i$~\cite{ci2008}. 
To choose between
the two (and to compute the $\ell_i$ component), we have to compute
$\lcp(i,\NSVT[i])$ and $\lcp(i,\PSVT[i])$ and choose the larger of the
two, see Fig.~\ref{fig-parsing-pseudo}.

\begin{figure}[th]
\centering
\begin{tabbing}
00:\=\qquad\=\qquad\=\qquad\=\qquad\=\qquad\=\kill
 \balgorithm{LZ-Factor$(i,psv,nsv)$}\\
 1:\>\bif $\lcp(i,psv) > \lcp(i,nsv)$ \bthen\\
 2:\>\>$(p,\ell) \la (psv,\,\lcp(i,psv))$\\
 3:\>\belse\\ 
 4:\>\>$(p,\ell) \la (nsv,\,\lcp(i,nsv))$\\
 5:\>\bif $\ell=0$ \bthen $p=\X[i]$ \\
 6:\>\boutput factor $(p,\ell)$\\
 7:\>\breturn $i + \max(\ell,1)$
\end{tabbing}
\caption{The basic procedure for computing a phrase starting at a position~$i$
given $psv=\PSVT[i]$ and $nsv=\NSVT[i]$. The return value is the
starting position of the next phrase.}
\label{fig-parsing-pseudo}
\end{figure}

\paragraph{Lazy LZ Factorization.}
The fastest LZ factorization algorithms in practice are from
recent papers by Kempa and Puglisi~\cite{kp2013} and Goto and
Bannai~\cite{gb2012}.  A common feature between them is a lazy
evaluation of LCP values: $\lcp(i,\NSVT[i])$ and $\lcp(i,\PSVT[i])$
are computed only when $i$ is a starting position of a phrase.  The
values are computed by a plain character-by-character comparison of
the suffixes, but it is easy to see that the total time complexity is
$\O(n)$.  This is in contrast to most previous algorithms that compute
the LCP values for every suffix using more complicated techniques.
The new algorithms in this paper use lazy evaluation too.

Goto and Bannai~\cite{gb2012} describe algorithms that compute and
store the full set of NSV/PSV values. One of their algorithms, BGT,
computes the $\NSVT$ and $\PSVT$ arrays with the help of the $\Phi$
array. The LZ factorization is then easily computed by repeatedly calling
\texttt{LZ-Factor}.  Two other algorithms, BGS and BGL, compute the
$\NSVL$ and $\PSVL$ arrays and use them together with $\SA$ and $\ISA$
to simulate $\NSVT$ and $\PSVT$ as in Eqs.~(\ref{eqn-NSVT})
and~(\ref{eqn-PSVT}). All three algorithms run in linear time and they
use $3n\log n$ (BGT), $4n\log n$ (BGL) and $(4n+s)\log n$ (BGS) bits
of working space, where $s$ is the size of the stack used by BGS.
In the worst case $s=\Theta(n)$.
The algorithms for computing the NSV/PSV values
are not new but come from~\cite{og2011} (BGT) and from~\cite{ci2008}
(BGL and BGS). However, the use of
lazy LCP evaluation makes the algorithms of Goto and Bannai faster
in practice than earlier algorithms. 

Kempa and Puglisi~\cite{kp2013} extend the lazy evaluation to the
NSV/PSV values too. Using $\ISA$ and a small data structure that
allows arbitrary NSV/PSV queries over $\SA$ to be answered quickly,
they compute $\NSVT[i]$ and $\PSVT[i]$ only when $i$ is a starting
position of a phrase. 
The approach requires $(2+1/b)n\log n$ bits of working space and 
$\O(n + zb + z\log(n/b))$ time, where $b$
is a parameter controlling a space-time tradeoff in the NSV/PSV data 
structure. If we set $b = \log n$, and given $z = \O(n/\log_{\sigma} n)$, 
then in the worstcase the algorithm requires $\O(n\log\sigma)$ time,
and $2n\log n+n$ bits of space. Despite the superlinear time complexity, this
algorithm (ISA9) is both faster and more space efficient than earlier
linear time algorithms.
Kempa and Puglisi also show how to reduce the space to $(1+\epsilon)n\log
n + n + \O(\sigma\log n)$ bits by storing a succinct representation 
of $\ISA$ (algorithms
ISA6r and ISA6s). Because of the lazy evaluation, these algorithms are
especially fast when the resulting LZ factorization is small.

\section{$3n\log n$-Bit Algorithm}

Our first algorithm is closely related to the algorithms of 
Goto and Bannai~\cite{gb2012}, particularly BGT and BGS. 
It first computes 
the $\PSVT$ and $\NSVT$ arrays and uses them for lazy LZ factorization 
similarly to
the BGT algorithm.  However, the NSV/PSV values are computed using the
technique of the BGS algorithm, which comes originally
from~\cite{ci2008}.  The algorithm is given
Figure~\ref{fig-13n-pseudo}.

\begin{figure}
\centering
\begin{tabbing}
00: \=\qquad\=\qquad\=\qquad\=\qquad\=\qquad\=\kill
 \balgorithm{KKP3}\\
 1:\>$\SA[0] \la 0$ \qquad // bottom of stack \\
 2:\>$\SA[n+1] \la 0$ \quad // empties the stack at end \\
 3:\>$top \la 0$ \qquad // top of stack\\
 4:\>\bfor $i\la 1$ \bto $n+1$ \bdo \\
 5:\>\>\bwhile $\SA[top] > \SA[i]$ \bdo \\
 6:\>\>\>$\NSVT[\SA[top]] \la \SA[i]$ \\
 7:\>\>\>$\PSVT[\SA[top]] \la \SA[top-1]$ \\
 8:\>\>\>$top \la top-1$ \qquad // pop from stack \\
 9:\>\>$top \la top+1$ \\
10:\>\>$\SA[top] \la \SA[i]$ \qquad // push to stack \\
11:\>$i \la 1$\\
12:\>\bwhile $i \le n$ \bdo\\
13:\>\>$i \la \texttt{LZ-Factor}(i,\PSVT[i],\NSVT[i])$
\end{tabbing}
\caption{LZ factorization using $3n \log n$ bits of working space (the
arrays $\SA$, $\NSVT$ and $\PSVT$).}
\label{fig-13n-pseudo}
\end{figure}

\newpage

The advantages of our algorithm compared to those of Goto and Bannai
are:
\begin{enumerate}
\item All of the algorithms of Goto and Bannai use an auxiliary array
  of size $n$, either $\ISA$ or $\Phi$. We need
  no such auxiliary array, which saves both space and time.
\item Both BGS and our algorithm need a stack whose maximum 
  size is not known in advance and can be $\Theta(n)$ in the worst
  case. BGS uses a dynamically growing separate stack while we
  overwrite the suffix array with the stack. This is possible because
  our algorithm makes just one pass over the suffix array (like BGT
  but unlike BGS) and the stack is never larger than the already
  scanned part of $\SA$.
\item Similar to the algorithms of Goto and Bannai, we store the
  arrays $\PSVT$ and $\NSVT$ interleaved so that the values $\PSVT[i]$
  and $\NSVT[i]$ are next to each other.  We compute the PSV value
  when popping from the stack instead of when pushing to the stack as BGS
  does. This way $\PSVT[i]$ and $\NSVT[i]$ are computed and written at
  the same time which can reduce the number of cache misses.
\end{enumerate}

\section{$\mathbf 2n\log n$-Bit Algorithm}

Our second algorithm reduces space by computing and storing only the
NSV values at first. It then computes the PSV values from the NSV
values on the fly.  As a side effect, the algorithm also computes the
$\Phi$ array!

For $t\in[0..n]$, let $\mathcal{X}_t=\{ \X[i..n] \mid i\le t\}$ be the
set of suffixes starting at or before position $t$. Let $\Phi_t$ be
$\Phi$ restricted to $\mathcal{X}_t$, that is, for $i\in[1..t]$,
suffix $\Phi_t[i]$ is the immediate lexicographical predecessor of
suffix $i$ among the suffixes in $\mathcal{X}_t$. In particular,
$\Phi_n=\Phi$.  As with the full $\Phi$, we make $\Phi_t$ a complete
unicyclic permutation by setting $\Phi_t[i_{\text{min}}]=0$ and
$\Phi_t[0]=i_{\text{max}}$, where $i_{\text{min}}$ and
$i_{\text{max}}$ are the lexicographically smallest and largest
suffixes in $\mathcal{X}_t$.  We also set $\Phi_0[0]=0$.  A~useful way
to view $\Phi_t$ is as a circular linked list storing $\mathcal{X}_t$
in the descending lexicographical order with $\Phi_t[0]$ as the head
of the list.

Now consider computing $\Phi_{t}$ given $\Phi_{t-1}$. 
We need to insert a new suffix
$t$ into the list, which can be done using standard insertion into a 
singly-linked list provided we know the position. It is easy to see
that $t$ should be inserted between $\NSVT[t]$ and $\PSVT[t]$. Thus
\begin{align*}
  \Phi_{t}[i] &= \left\{
    \begin{array}{ll}
      t & \text{ if } i = \NSVT[t] \\
      \PSVT[t] & \text{ if } i=t \\
      \Phi_{t-1}[i] & \text{ otherwise}
    \end{array}
  \right.
\end{align*}
and furthermore
\begin{align*}
  \PSVT[t] &= \Phi_{t-1}[\NSVT[t]] \ .
\end{align*}

The pseudocode for the algorithm is given in
Figure~\ref{fig-9s-pseudo}. The NSV values are computed essentially
the same way as in the first algorithm (lines~1--9) and stored in the
array $\Phi$.  In the second phase, the algorithm maintains the
invariant that after $t$ rounds of the loop on lines~12--18, 
$\Phi[0..t]=\Phi_t$ and $\Phi[t+1..n]=\NSVT[t+1..n]$.

\begin{figure}
\centering
\begin{tabbing}
00: \=\qquad\=\qquad\=\qquad\=\qquad\=\qquad\=\kill
 \balgorithm{KKP2s}\\
 1:\>$\SA[0] \la 0$ \qquad // bottom of stack \\
 2:\>$\SA[n+1] \la 0$ \quad // empties the stack at end \\
 3:\>$top \la 0$ \qquad // top of stack\\
 4:\>\bfor $i\la 1$ \bto $n+1$ \bdo \\
 5:\>\>\bwhile $\SA[top] > \SA[i]$ \bdo \\
 6:\>\>\>$\Phi[\SA[top]] \la \SA[i]$ \quad // $\Phi[\SA[top]] = \NSVT[\SA[top]]$\\
 7:\>\>\>$top \la top-1$  \qquad // pop from stack \\
 8:\>\>$top \la top+1$ \\
 9:\>\>$\SA[top] \la \SA[i]$  \qquad // push to stack \\
10:\>$\Phi[0] \la 0$\\
11:\>$next \la 1$\\
12:\>\bfor $t \la 1$ \bto $n$ \bdo \\
13:\>\>$nsv \la \Phi[t]$\\
14:\>\>$psv \la \Phi[nsv]$\\
15:\>\>\bif $t=next$ \bthen\\
16:\>\>\>$next \la \texttt{LZ-Factor}(t,psv,nsv)$\\
17:\>\>$\Phi[t] \la psv$\\
18:\>\>$\Phi[nsv] \la t$
\end{tabbing}
\caption{LZ factorization using $2n \log n$ bits of working space (the
arrays $\SA$ and $\Phi$).}
\label{fig-9s-pseudo}
\end{figure}

\section{Getting Rid of the Stack}

The above algorithms overwrite the suffix array with the stack, which
can be undesirable. First, we might need the suffix array later for
another purpose. Second, since the algorithms make just one sequential
pass over the suffix array, we could stream the suffix array from disk
to further reduce the memory usage.  In this section, we describe
variants of our algorithms that do not overwrite $\SA$ (and still make
just one pass over it).

The idea is to replace the stack with $\PSVT$ pointers. If $j$ is the
suffix on the top of the stack, then the next suffixes in the stack are
$\PSVT[j]$, $\PSVT[\PSVT[j]]$, etcetera. This can be easily seen in
how the $\PSVT$ values are computed in KKP3 (line 7 in
Fig.~\ref{fig-13n-pseudo}). Thus given $\PSVT$ we do not need an explicit
stack at all. Both of our algorithms can be modified to exploit this:
\begin{itemize}
\item In KKP3, we need to compute the $\PSVT$ values when pushing on
  the stack rather than when popping. The body of the main loop (lines 5--10
  in Fig.~\ref{fig-13n-pseudo})
  now becomes:
\begin{tabbing}
00: \=\qquad\=\qquad\=\qquad\=\qquad\=\qquad\=\kill
\>\>\bwhile $top > \SA[i]$ \bdo\\
\>\>\>$\NSVT[top] \la \SA[i]$\\
\>\>\>$top \la \PSVT[top]$\\
\>\>$\PSVT[\SA[i]] \la top$\\
\>\>$top \la \SA[i]$
\end{tabbing}
\item KKP2s needs to be modified to compute $\PSVT$ values first instead of
  $\NSVT$ values. The $\PSVT$-first version is symmetric to the
  $\NSVT$-first algorithm. In particular, $\Phi_t$ is replaced by
  the inverse permutation $\Phi_t^{-1}$.
  The algorithm is shown in
  Fig.~\ref{fig-9n-pseudo}.
\end{itemize}

\begin{figure}
\centering
\begin{tabbing}
00: \=\qquad\=\qquad\=\qquad\=\qquad\=\qquad\=\kill
 \balgorithm{KKP2n}\\
 1:\>$top \la 0$ \qquad // top of stack\\
 2:\>\bfor $i\la 1$ \bto $n$ \bdo \\
 3:\>\>\bwhile $top > \SA[i]$ \bdo \\
 4:\>\>\>$top \la \Phi^{-1}[top]$  \quad // pop from stack \\
 5:\>\>$\Phi^{-1}[\SA[i]] \la top$ \quad // $\Phi^{-1}[\SA[i]] = \PSVT[\SA[i]]$\\
 6:\>\>$top \la SA[i]$ \qquad // push to stack \\
 7:\>$\Phi^{-1}[0] \la 0$\\
 8:\>$next \la 1$\\
 9:\>\bfor $t \la 1$ \bto $n$ \bdo \\
10:\>\>$psv \la \Phi^{-1}[t]$\\
11:\>\>$nsv \la \Phi^{-1}[psv]$\\
12:\>\>\bif $t=next$ \bthen\\
13:\>\>\>$next \la \texttt{LZ-Factor}(t,psv,nsv)$\\
14:\>\>$\Phi^{-1}[t] \la nsv$\\
15:\>\>$\Phi^{-1}[psv] \la t$
\end{tabbing}
\caption{LZ factorization using $2n \log n$ bits of working space (the
arrays $\SA$ and $\Phi^{-1}$) without an explicit stack. The $\SA$ remains intact
after the computation.}
\label{fig-9n-pseudo}
\end{figure}

The versions without an explicit stack are slightly slower because of
the non-locality of the pointer accesses. If we need to avoid
overwriting $\SA$, a faster alternative would be to use a separate
stack. However, the stack can grow as big as $n$ (for example when
$\X=a^{n-1}b$) which increases the worst case space requirement by
$n\log n$ bits.

We can get the best of both alternatives by adding a fixed size
stack buffer to the stackless version. The buffer holds the top part
of the stack to speed up stack operations.  When the buffer gets full,
the bottom half of its contents is discarded, and when the buffer gets empty,
it is filled half way using the PSV pointers. This version is called
KKP2b. The time complexity remains linear and is independent of the buffer size.

\section{Experimental Results} \label{sec:experiments}

We implemented the algorithms described in this paper and compared their performance 
in practice to algorithms from~\cite{kp2013} and~\cite{gb2012}.
Experiments measured the time to compute the LZ factorization of the text.
All algorithms take the text and the suffix array as an input hence we omit the
time to compute $\SA$. The data set used in experiments is described in detail in
Table~\ref{tab-files}.

\begin{table*}
  \centering {
  \begin{tabular}[tab:space-basic]{l@{\hspace{1.9em}}c@{\hspace{1.9em}}r@{\hspace{1.9em}}r@{\hspace{1.9em}}l@{\hspace{1.9em}}l}
\hline
Name & $n/2^{20^{\vphantom{a}}}$ & $\sigma$ & $n/z$ & Source & Description \\
\hline
proteins        & 150  & 25     & 9.57  & S    & Swissprot database\\
english         & 150 & 220     & 13.77 & S    & Gutenberg Project\\
dna             & 150 & 16      & 14.65 & S    & Human genome\\
sources         & 150 & 228     & 17.67 & S    & Linux and GCC sources\\
\hline
coreutils       & 150 & 236     & 110   & R/R  & 9 $\times$ GNU Coreutils source\\
cere            & 150 & 5       & 112   & R/R  & 37 $\times$ baking yeast genome\\
kernel          & 150 & 160     & 214   & R/R  & 36 $\times$ Linux Kernel sources\\
einstein.en     & 150 & 124     & 3634  & R/R  & Wikipedia\\
\hline
proteins.001.1  & 100 & 21      & 295   & R/PR & 100 $\times$ 1MB protein \\
english.001.2   & 100 & 106     & 312   & R/PR & 100 $\times$ 1MB english \\
dna.001.1       & 100 & 5       & 340   & R/PR & 100 $\times$ 1MB dna \\
sources.001.2   & 100 & 98      & 355   & R/PR & 100 $\times$ 1MB sources \\
\hline
tm29            & 150 & 2       & 2912K & R/A  & Thue-Morse sequence\\
rs.13           & 150 & 2       & 3024K & R/A  & Run-Rich String sequence\\
\hline
  \end{tabular} }\vspace{1ex}
\caption[lof]{Files used in the experiments. The files are from the standard (S)
  Pizza\&Chili corpus (\mbox{\url{http://pizzachili.dcc.uchile.cl/texts.html}})
  and from the repetitive (R) Pizza\&Chili corpus
  (\mbox{\protect\url{http://pizzachili.dcc.uchile.cl/repcorpus.html}}).
  The repetitive corpus consists of files containing multiple copies of similar
  data (R), artificially generated sequences (A), and files created from standard
  corpus by concatenating 100 copies of 1MB prefix and mutating them randomly
  (PR). The value of $n/z$ (average length of phrase in LZ factorization) is
  included as measure of repetitiveness.}
  \label{tab-files}
\end{table*}

\paragraph{Experiments Setup.}
We performed experiments on a 2.4GHz Intel Core i5 CPU equipped with 3072KB L2
cache and 4GB of main memory. 
The machine had no other significant CPU tasks running and  
only a single thread of execution was used. 
The OS was Linux (Ubuntu 10.04, 64bit) running kernel 2.6.32. All programs were compiled using
{\tt g++} version 4.4.3 with {\tt-O3} {\tt-static} {\tt-DNDEBUG} options. For
each combination of algorithm and test file we report the median runtime from 
five executions. The times were recorded with the standard C {\tt clock}
function. All data structures reside in main memory during computation.

\paragraph{Discussion.}
In nearly all cases algorithms introduced in this paper outperform the algorithms
from~\cite{gb2012} (which are, to our knowledge, the fastest up-to-date linear time LZ
factorization algorithms) while using the same or less space. In particular the
KKP2 algorithms are always faster and simultaneously use at least $n \log n$ bits
less space. A notably big difference is observed for non-repetitive data, where
KPP3 significantly dominates all prior solutions.

The new algorithms (e.g. KKP2b) also dominate in most cases the general purpose practical
algorithms from~\cite{kp2013} (ISA9 and ISA6s), while offering stronger worst
case time guarantees, but are a frame slower (and use about 50\% more space in
practice) than ISA6r for highly repetitive data.

The comparison of KKP2n to KKP2s reveals the expected slowdown (up to 16\%) 
due to the non-local stack simulation. However, this effect is almost
completely eliminated by buffering the top part of the stack (KKP2b). With a
256KB buffer we obtained runtimes almost identical to KKP2s ($<1\%$
difference in all cases). We observed a similar effect when applying this 
optimization to the KKP3
algorithm but, for brevity, we only present the improvement for KKP2.

Finally, we observe that KKP2b, despite being slower than KPP3 on non-repetitive
data, has runtimes that are very close to best in each category,
making it perhaps the most applicable general purpose algorithm.

\begin{table*}
  \centering {
  \begin{tabular}[tab:space-basic]{l@{\hspace{0.7em}}|@{\hspace{0.5em}}c@{\hspace{0.2em}}c@{\hspace{0.2em}}c@{\hspace{0.2em}}c@{\hspace{0.4em}}|@{\hspace{0.6em}}c@{\hspace{0.9em}}c@{\hspace{0.9em}}c@{\hspace{0.7em}}|@{\hspace{0.7em}}c@{\hspace{0.9em}}c@{\hspace{0.9em}}c}
\hline
Testfile         & KKP3    & KKP2s    & KKP2b    & KKP2n    & ISA6r   & ISA6s   & ISA9    & iBGS   & iBGL    & iBGT\\
\hline
proteins         & 74.5  & 83.9  & 84.1  & 88.1    &    -    & 198.0  & 92.7   & 99.8  & 123.2  & 171.4\\
english          & 75.7  & 80.6  & 80.6  & 84.6    &    -    & 171.0  & 83.9   & 93.2  & 108.6  & 153.9\\
dna              & 81.7  & 92.7  & 92.7  & 97.3    &    -    & 175.2  & 86.1   & 97.5  & 113.4  & 188.0\\
sources          & 50.5  & 54.7  & 54.8  & 56.1    &    -    & 115.0  & 59.3   & 69.3  & 77.8   & 99.8\\
\hline
coreutils        & 43.6  & 40.2  & 40.2  & 40.6    &  43.3  & 49.4   & 41.9   & 51.5  & 52.2   & 55.4\\
cere             & 63.2  & 53.3  & 53.2  & 57.7    &  51.8  & 56.3   & 53.0   & 65.5  & 66.1   & 84.1\\
kernel           & 45.7  & 41.6  & 41.5  & 42.2    &  39.2  & 45.7   & 42.8   & 52.9  & 53.0   & 56.2\\
einstein.en      & 56.9  & 43.6  & 43.5  & 47.6    &  31.1  & 37.1   & 45.2   & 60.0  & 58.6   & 52.8\\
\hline
proteins.001.1   & 52.6  & 43.1  & 43.1  & 50.0    &  40.7  & 45.3   & 46.6   & 58.4  & 57.6   & 59.6\\
english.001.2    & 52.0  & 43.4  & 43.8  & 52.2    &  40.4  & 45.1   & 45.3   & 57.7  & 56.0   & 79.4\\
dna.001.1        & 55.6  & 43.9  & 43.9  & 50.8    &  39.2  & 43.7   & 45.0   & 58.5  & 57.8   & 62.5\\
sources.001.2    & 43.1  & 40.5  & 40.5  & 47.8    &  38.0  & 42.8   & 41.1   & 54.2  & 52.4   & 72.2\\
\hline
tm29             & 38.2  & 35.1  & 35.1  & 38.7    &  34.2  & 39.6   & 36.4   & 44.1  & 44.2   & 44.4\\
rs.13            & 77.8  & 49.0  & 49.4  & 52.0    &  34.8  & 40.8   & 51.8   & 59.5  & 59.5   & 56.5\\
\hline
  \end{tabular} }\vspace{1ex}
\caption{Times for computing LZ factorization. The times are seconds per gigabyte and do not
include any reading from or writing to disk.}
\label{tab-times}
\end{table*}

\section{Future Work} \label{sec:conclusions}

For data of low to medium repetitiveness the algorithms introduced in this paper 
are the fastest available. These algorithms should adapt easily to a semi-external
setting because, apart from the need to permute the NSV/PSV values into text order, which 
can be handled in external memory, all non-sequential memory accesses are restricted 
to the input string. We are currently exploring this direction.

There are several interesting open problems. One is the need for a fully external
memory algorithm for LZ factorization, especially given the recent pattern matching indexes 
which use LZ77. Relatedly, parallel and 
distributed approaches are also of high interest. A recent step in the external memory 
direction is~\cite{kp2013b}. 

Another problem is to find a scalable way to accurately 
estimate the size of the LZ factorization in lieu of actually computing it. Such a 
tool would be useful for entropy estimation, and to guide the selection of appropriate 
compressors and compressed indexes when managing massive data sets. 

Finally, one wonders if only $(1+\epsilon)n \log n + \O(\log n)$ bits of working memory is enough for 
linear runtime. The most space-efficient algorithm in this paper use $2n\log n + \O(\log n)$ 
bits, and in~\cite{kp2013} working space of $(1 + \epsilon)n\log n + n + \O(\sigma\log n)$ bits 
(for arbitrary constant $\epsilon$) is achieved, but at the price of $\O(n\log\sigma)$ runtime.

\section*{Acknowledgments}

Our thanks go to Keisuke Goto and Hideo Bannai for sending us a
preliminary copy of their manuscript.

\bibliographystyle{plain}
\bibliography{lz}

\end{document}